\documentstyle[12pt,aj_pt4]{article}

\lefthead{Nitta et al.}
\righthead{Self-similar Evolution of Fast Reconnection}

\begin{document}

\title{Fast magnetic reconnection in free space: self-similar evolution process}

\author{S. Nitta \altaffilmark{1}}
\affil{Department of Astronomical Science, The Graduate University for Advanced Studies, 2-21-1 Osawa, Mitaka 181-8588, Japan}
\authoremail{snitta@th.nao.ac.jp}

\author{S. Tanuma \altaffilmark{2}$^,$\altaffilmark{3}}
\affil{Department of Astronomy, School of Science, University of Tokyo,
7-3-1 Hongo, Bunkyo-ku, Tokyo 113-0033, Japan}
\authoremail{tanuma@solar.mtk.nao.ac.jp}

\author{K. Shibata}
\affil{Kwasan and Hida Observatories, Kyoto University, Yamashina, Kyoto 607-8417, Japan}
\authoremail{shibata@kwasan.kyoto-u.ac.jp}

\and

\author{K. Maezawa}
\affil{Institute of Space and Astronautical Science 3-1-1 Yoshinodai, Sagamihara 229-8510 Japan}
\authoremail{maezawa@stp.isas.ac.jp}

\altaffiltext{1}{Division of Theoretical Astrophysics, National Astronomical Observatory of Japan, 2-21-1 Osawa , Mitaka 181-8588, Japan}

\altaffiltext{2}{Division of Solar Physics, National Astronomical Observatory of Japan, 2-21-1 Osawa, Mitaka 181-8588, Japan}

\altaffiltext{3}{Current address: Solar-Terrestrial Environment Laboratory, Nagoya University, 3-13 Honohara, Toyokawa, Aichi 442-8507, Japan}

\begin{abstract}
We present a new model for time evolution of fast magnetic reconnection in free space, which is characterized by self-similarity. Reconnection triggered by locally enhanced resistivity assumed at the center of the current sheet can self-similarly and unlimitedly evolve until external factors affect the evolution. The possibility and stability of this type of evolution are verified by numerical simulations in a very wide spatial dynamic range. Actual astrophysical reconnection in solar flares and geomagnetospheric substorms can be treated as an evolutionary process in free space, because the resultant scale is much larger than the initial scale. In spite of this fact, most of the previous numerical works focused on the evolutionary characters strongly affected by artificial boundary conditions on the simulation boundary. Our new model clarifies a realistic evolution for such cases. The characteristic structure around the diffusion region is quite similar to the Petschek model which is characterized by a pair of slow-mode shocks and the fast-mode rarefaction-dominated inflow. However, in the outer region, a vortex-like return flow driven by the fast-mode compression caused by the piston effect of the plasmoid takes place. The entire reconnection system expands self-similarly.
\end{abstract}


\section{Introduction}
\label{sec:Int}

It is widely accepted that magnetic reconnection takes place very commonly as a very powerful energy converter in astrophysical plasma systems. However, there still remain so many open questions not only for the microscopic physics of the resistivity, but also for the macroscopic magnetohydrodynamical (MHD) structure. 

One of the most important questions on magnetic reconnection is what condition determines the reconnection rate, by which we can estimate the speed of energy conversion. The reconnection rate is usually defined as the ratio of the Alfv\'{e}n wave transit time scale across the system length to the energy conversion time scale. The reconnection rate mainly depends upon the magnetic Reynolds number $R_m$ which represents the ratio of the diffusion time scale of the magnetic field to the Alfv\'{e}n wave transit time scale. We should note that the magnetic diffusivity of astrophysical plasma systems is very small ($R_m \sim 10^{14}$ for the solar corona). We here review the previous models of magnetic reconnection from the viewpoint of the reconnection rate. 

Models in which the magnetic energy is mainly converted by magnetic diffusion (e.g., the Sweet-Parker model [Sweet 1958, Parker 1963]) are called ``slow'' reconnection (Tajima \& Shibata 1997), because these models crucially depend upon $R_m$ (i.e., the reconnection rate of the Sweet-Parker model is in proportion to ${R_m}^{-1/2}$). Hence, the reconnection rates of these models are very small, and these models cannot convert the magnetic energy in a short period (e.g., the typical duration of solar flares is $\sim$ a few minutes-a few hours). These slow reconnection models cannot provide adequate reconnection rates for actual powerful energy conversion. 

A reconnection model with a pair of slow-mode shocks is firstly presented by Petschek (1964). In this model, the magnetic energy is mainly converted at the slow-mode shocks. Therefore, the time scale of the energy conversion is determined by MHD wave propagation, and is much shorter than that of the simple diffusion model or of the Sweet-Parker model. The reconnection rate is almost independent of the magnetic Reynolds number ($\propto \log R_m \sim {R_m}^0$). Hence, we call this quick magnetic energy converter ``fast reconnection''. 

We should note, however, that the question of how we can attain the fast reconnection is still open. Concerning this point, there are two different models. One states that external boundary conditions should control the reconnection, so that the resistivity has no effective influence on the energy conversion (see, e.g., Petschek 1964, Priest \& Forbes 1986 for theoretical studies, and Sato \& Hayashi 1979 for numerical studies). The other has been proposed in a series of numerical works originated by Ugai \& Tsuda (Ugai \& Tsuda 1977, 1979, Tsuda \& Ugai 1977, and recently, Ugai 1999). They put a localized finite resistivity in the current sheet assuming that such resistivity simulates the anomalous resistivity in plasmas. This resistivity acts as a trigger to start the magnetic reconnection. Once this resistivity is put, the reconnection starts and evolves self-consistently, and forms a fast reconnection system with the Petschek-type structure (a pair of slow-mode shocks is formed along the current sheet). 

However, even in the previous numerical studies aimed at clarification of the time evolution of fast reconnection, the evolution could not be followed for a long time. This is mainly due to the finite size of the simulation box, hence the application of the results was limited to spatial scales typically, say, hundred times the spatial scale of the diffusion region. 

The actual magnetic reconnection in astrophysical systems usually grows in a huge dynamic range in its spatial dimension. For example, the initial scale of the reconnection system can be defined by the initial current sheet thickness, but this is too small to be observed in typical solar flares. We do not have any convincing estimate of it, but if we estimate it to be of the order of the ion Larmor radius, it is so small ($\sim 10^0$ m in the solar corona). Finally, the reconnection system develops on the scale of the order of the initial curvature radius of the magnetic filed lines ($\sim 10^7$ m $\sim$ 1.5\% of the solar radius for typical solar flares). The dynamic range of the spatial scale is obviously huge ($\sim 10^7$ for solar flares). For the geomagnetospheric substorms, their dynamic range of growth is also large ($\sim 10^4$ for substorms). Such a very wide dynamic range of growth suggests that the evolution of the magnetic reconnection should be treated as a development in ``free space'', and boundary conditions do not affect the evolutionary process of magnetic reconnection, at least in the initial stage just after the onset of reconnection. 

In the actual numerical studies, due to the restriction of computer memory, the dimension of the simulation box is limited.  For magnetic reconnection to be properly studied, the thickness of the current sheet must be sufficiently resolved by simulation mesh size. Usually, the thickness of the current sheet is much smaller than the size of the entire system. Hence, we are forced to cut a finite volume out from actual large-scale reconnection system for numerical studies. 

Of course, we wish that the boundary conditions of this type of finite simulation box reproduces the evolution in unbounded space. However, we should note that, in actual simulations, the boundary of such a simulation box necessarily affects the evolution inside the box, even if we take so-called ``free'' boundary conditions. This fact has an adverse influence on the evolutionary process of the numerically simulated magnetic reconnection as follows. When the reconnection proceeds and physical signals propagating from the inner region cross the boundaries of the simulation box, the subsequent evolution is necessarily affected by the boundary conditions. This is because the information propagated through the boundary is completely lost and such an artificially cut-out simulation box will never receive the proper response of the outer regions. Additionally, numerical and unphysical signals which are emitted from the boundary may disturb the evolution. Such artificially affected evolution is obviously unnatural, and the resultant stationary state should be different from the actual reconnection occurring in free space. 

In this paper, our interest will be focused on the evolutionary processes of magnetic reconnection in ``free space''. On the basis of numerical simulations, we propose a new model for time-dependent magnetic reconnection. In this model, an entire system of magnetic reconnection triggered by locally enhanced resistivity continues to evolve self-similarly and unlimitedly. We call our model the self-similar evolution model. Propagation of the fast-mode rarefaction wave emitted from the reconnection region plays crucial roles in this self-similar evolution model. This point has never been pointed out in the previous works. The launch of rarefaction waves induces a flow toward the current sheet, which leads to a characteristic inflow structure. We suggest that this characteristic structure is related to the observational results on solar flares (see section \ref{sec:App}). Our reconnection system is accompanied by a pair of slow-mode shocks caused by the induced inflow just as in the Petschek model. Majority of the magnetic energy is converted at this slow shock. Hence, this is categorized as fast reconnection. The scenario for the evolutionary process is clarified in section \ref{sec:Sce}, and properties of this new model are discussed in section \ref{sec:S-D}. The results of numerical simulations performed in this paper are summarized in section \ref{sec:Num}. On the basis of such numerical studies, we obtain an intuition to the model. This kind of intuition will stimulate our theoretical effort to find an analytic representation of the self-similar evolution model.

\section{Scenario of Evolutionary Process}
\label{sec:Sce}

The scenario of the self-similar evolution of magnetic reconnection is summarized as follows. This paper extends in a greater detail some of the themes already sketched in Nitta (1988). 

We suppose a global MHD equilibrium for an anti-parallel magnetic field configuration with an embedded current sheet as the initial state. We apply Cartesian coordinates, whose $x$-axis is taken along the current sheet, and $y$-axis perpendicular to the current sheet. We assume uniformity in the $z$-direction and hence treat a two-dimensional MHD problem. Hereafter we study only the configuration in the $x-y$ plane. 

The initial state is supposed to be the Harris solution (see figure \ref{fig:sim_boxf1}), 
\begin{equation}
B_x=B_0 \tanh (y/D) \ , 
\end{equation}
\begin{equation}
P=\frac{{B_0}^2}{8\pi}\left(\beta+\frac{1}{\cosh^2 (y/D)}\right) \ ,
\end{equation}
where $D$ is the initial thickness of the current sheet, $B_0$ the asymptotic strength of magnetic field, $P$ the gas pressure, and $\beta$ the ratio of gas to magnetic pressure (so-called plasma $\beta$ value) in the asymptotic region ($|y| \gg D$). We should remark again that our attention is focused on the free evolution without any influence of the boundary conditions.

\placefigure{fig:sim_boxf1}

\noindent
1) Onset of reconnection (resistive stage: $t \mathrel{\hbox{\rlap{\hbox{\lower4pt\hbox{$\sim$}}}\hbox{$<$}}} D/V_{A0}$)\\

When a microscopic disturbance takes place, a localized resistivity is enhanced in the current sheet (see figure \ref{fig:sim_boxf1}). This resistivity induces magnetic diffusion, and anti-parallel magnetic field lines begin to reconnect. This stage will correspond to the Sweet-Parker type reconnection or the resistive tearing mode instability. Once the reconnection of field lines begins, the total pressure in the vicinity of the reconnection point should decrease due to the ejection of a pair of bipolar plasma jets (the reconnection jets) along the current sheet (see figure \ref{fig:stage1f2}). 

In a low-$\beta$ plasma ($\beta \ll 1$ in the asymptotic region; this is the usual case in astrophysical problems), the propagation speed of the fast-magnetosonic wave is much larger than that of other wave modes. The information of decreasing total-pressure near the reconnection point propagates almost isotropically as a fast-mode rarefaction wave (hereafter we call it FRW) with a speed almost equal to the Alfv\'{e}n speed $V_{A0}$ in the asymptotic region ($|y| \gg D$). Hence the wave front of FRW (hereafter we call it FRWF) has a circular shape except near the point where FRWF touches the current sheet. \\

\noindent
2)Induction of the inflow ($t > D/V_{A0}$)\\

As FRW propagates in the asymptotic region, the resultant total-pressure gradient induces a plasma inflow toward the current sheet (see figure \ref{fig:stage2f3}). Outside FRWF, plasmas never move because no signal reaches that region yet. Hence the expansion speed and the shape of FRWF are kept constant throughout the evolution. \\

\noindent
3)Self-similar evolution (self-similarly evolving fast reconnection stage: $t \gg D/V_{A0}$)\\

When the flow toward the current sheet develops sufficiently, a pair of slow-mode shocks forms along the current sheet. After formation of the slow shock, the energy conversion is drastically enhanced. The majority of the magnetic energy is converted at the slow shock. Hence this stage represents ``fast reconnection''. Once this system of fast reconnection is set up, the dimension of the system unlimitedly develops self-similarly (see figure \ref{fig:stage3f4}). \\

This scenario of the evolution is confirmed by our numerical simulations. The outline and the results of the simulations are shown in the next section.

\placefigure{fig:stage1f2}

\placefigure{fig:stage2f3}

\placefigure{fig:stage3f4}

\section{Numerical Simulations}
\label{sec:Num}

The evolution of magnetic reconnection in free space is numerically studied in the following subsections.

\subsection{Outlines}
\label{sec:Out}

We simulate the time evolution of the 2-D magnetic reconnection in free space (see figure \ref{fig:sim_boxf1}). The initial equilibrium state is supposed to be expressed by the Harris solution (see section \ref{sec:Sce}). We impose symmetry conditions on the lines $x=0$ and $y=0$, and find solution for the quadrant $0 \leq x$ and $0 \leq y$. A locally enhanced resistivity is put at the center of the current sheet (i.e., in the region $0 \leq x \leq 2D$, $0 \leq y \leq 2D$ in figure \ref{fig:sim_boxf1}). In this region, the value of the resistivity is assumed to be uniform and is kept constant throughout the evolution. 

Our attention is focused on such evolutionary processes that are free from the influence of outer boundary conditions. For this purpose, we place the outer boundaries as far from the reconnection region as possible, and study the evolution before any signal from the central region reaches the boundaries. To realize such a wide simulation region, we adopt a non-uniform mesh as follows. In the vicinity of the diffusion region ($x,y \leq 20 D$), the mesh size is $D/10$ in both the $x$ and $y$ directions to obtain a high spatial resolution. Outside this region, the mesh size increases exponentially in both $x$ and $y$ directions. The maximum mesh size is $30 D$ in both directions (as a result, our mesh has a very long, slender shape at large distances along the $x$ or $y$ axis). 

We use the 2-step Lax-Wendroff scheme with artificial viscosity to solve the following 2-D resistive MHD equations. 

\begin{eqnarray}
{\partial\rho\over\partial t}+\mbox{\boldmath $\nabla$}\cdot(\rho\mbox{\boldmath$v$})&=&0,\\
\rho{\partial\mbox{\boldmath$v$}\over\partial t}
+\rho(\mbox{\boldmath$v$}\cdot\mbox{\boldmath$\nabla$})\mbox{\boldmath$v$}
+\mbox{\boldmath$\nabla$} P
&=& {1\over c} \mbox{\boldmath$J$}\times\mbox{\boldmath$B$},\\
{\partial\mbox{\boldmath$B$}\over\partial t}
-\mbox{\boldmath$\nabla$}\times(\mbox{\boldmath$v$}\times\mbox{\boldmath$B$})
&=& -c \mbox{\boldmath$\nabla$}\times (\eta \mbox{\boldmath$J$}),\\
{\partial e\over\partial t}
+\mbox{\boldmath$\nabla$}\cdot\left[(e+P)\mbox{\boldmath$v$}\right]
&=& \eta |\mbox{\boldmath$J$}|^2+\mbox{\boldmath$v$} \cdot \mbox{\boldmath$\nabla$} P
\end{eqnarray}
where $\rho$, $\mbox{\boldmath$v$}$, $P$, $\mbox{\boldmath$B$}$, $\eta$, $e$ and $\mbox{\boldmath$J$}$ are the mass density, velocity, gas pressure, magnetic field, magnetic diffusivity, internal energy, and current density ($=c\mbox{\boldmath$\nabla$}\times\mbox{\boldmath$B$}/4\pi$), respectively. We use the equation of state for ideal gas,
i.e., $P=(\gamma-1)e$
where $\gamma$ is the specific heat ratio (=5/3).

\subsection{Parameters \& Normalization}
\label{sec:Par}

The only parameters of this simulation are the plasma $\beta$ value and the magnetic Reynolds number $R_m$. $\beta^{1/2}$ denotes the ratio of the Alfv\'{e}n wave transit time scale to the sound transit time scale. Similarly, $R_m$ denotes the ratio of the magnetic diffusion time scale to the Alfv\'{e}n wave transit time scale. In most cases of astrophysical problems, we can set $\beta$ to be much smaller than unity (e.g., $\beta \sim 10^{-2}$ for solar corona). $R_m$ is much larger than unity (e.g., $R_m \sim 10^{14}$ for solar flares). However, the situation with $\beta \ll 1$ and/or $R_m \gg 1$ is very difficult to treat numerically, because we must solve the problem with a very wide dynamic range of time scales. Therefore, we approximately choose moderate values for these parameters as follows (We call this as typical case or case A; see below). 

\begin{equation}
\beta \equiv \frac{P_0}{{B_0}^2/8 \pi} =0.2 \ ,
\end{equation}
\begin{equation}
R_m \equiv \frac{V_{A0}}{\eta/D} =24.5 \ ,
\end{equation}
where $P_0$, $B_0$ and $V_{A0}$ are the gas pressure, magnetic field strength and Alfv\'{e}n speed in the asymptotic region far from the neutral sheet (uniform in this region), respectively, and $\eta$ is the magnetic diffusivity in the diffusion region ($x, y \leq 2 D$, see figure \ref{fig:sim_boxf1}). 

We normalize physical quantities as follows: the constants $V_{A0}$, $D$ and $\rho_0$ are used to normalize the dimensions [L T$^{-1}$], [L] and [M L$^{-3}$], respectively, where $V_{A0}$ and $\rho_0$ are the Alfv\'{e}n speed and the mass density in the asymptotic region, respectively. In the normalized units, the values of other physical quantities are expressed as $P_0=0.6$, $B_0=8.7$, $c_{s0}=0.4$ and $\eta=0.1$, where $c_{s0}$ is the sound speed at the initial state (uniform in the simulation box). 

We have performed numerical simulations for the following models. The main results discussed in the following sections are based on case A. Other cases are performed for comparison to the typical case A. 

Case A: Typical case (figures \ref{fig:series1f5}, \ref{fig:pm_vf7}, \ref{fig:profile_rhof8}, \ref{fig:profile_pgf9}, \ref{fig:profile_bxf10}, \ref{fig:profile_vxf11}, \ref{fig:profile_vyf12}, \ref{fig:det-totf13}, \ref{fig:det-magf14}, \ref{fig:dimmingf16} and \ref{fig:rec_ratef17}). Details are explained in the previous section. 

Case A': Larger simulation box (figure \ref{fig:series2f6}). Mesh size is coarser than the typical case in order to study the evolution subsequent to case A. The maximum mesh size is 100 in both directions. Other parameters are identical to the typical case. We use the final state of the simulation for case A as the initial condition for this case. 

Case B: Circular resistive region (Nitta 1988). The resistivity is distributed in a circular region ($r \leq 1$ where $r$ is distance from the origin) as an exponentially decaying function of $r$ with maximum value at the center. 

Case C: Larger resistivity (figure \ref{fig:case_c_pmf15}). The resistivity  ($\eta =0.5$) is larger than the typical case ($\eta=0.1$). Other parameters are identical to the typical case. 

Case D: Unmagnetized upper half region (figure \ref{fig:rec_ratef17} [dashed line]). As the initial condition, the lower half region ($y \leq 3000$) is filled with uniformly magnetized plasma, while the upper half region ($y>3000$) is unmagnetized. The magnetic field falls off as a tangent-hyperbolic function of $y$ with the characteristic decaying scale of 1000. Other parameters are identical to the typical case.

\subsection{Simulation Results}
\label{sec:Sim}

The most significant feature of our results is the self-similar expansion of the entire reconnection system, which means that a solution at a particular moment is the same as that at past or in future if we change the spatial scaling in proportion to time from the onset. A sequence of six snapshots shown in figure \ref{fig:series1f5} shows the time evolution of the reconnection system. The subsequent evolution (case A') is shown in figure \ref{fig:series2f6} (To follow the evolution for a longer time, we adopted larger meshes than figure \ref{fig:series1f5}). Each figure represents a snapshot of the evolution at the denoted time. Note that the scale of the both axes expands in proportion to time. We call this ``zoom-out coordinates''. The color contours of the magnetic pressure and the configuration of the magnetic field lines are shown in figure \ref{fig:series1f5} and \ref{fig:series2f6}. The blue arc denotes FRWF. The sequence in figures \ref{fig:series1f5} and \ref{fig:series2f6} clearly shows that the distribution of physical quantities (e.g., the magnetic pressure and the magnetic field lines) approaches to a stationary solution in the zoom-out coordinates. As we will see, other quantities have the same property. 

In order to remove ambiguity of the term ``self-similar evolution'', let us consider the following situation. Let distribution of a physical quantity, say $Q$, to be time varying in the fixed coordinate (i.e., $Q=f[\mbox{\boldmath$r$}, t]$). When we measure the distribution of $Q$ in the zoom-out coordinate, if $Q=g(\mbox{\boldmath$r$}')$ (where $\mbox{\boldmath$r$}' \equiv \mbox{\boldmath$r$}/[V_{A0} t]$ is the position vector from the origin [the center of the diffusion region] in the zoom-out coordinate, with $\mbox{\boldmath$r$}$ the position vector in the original fixed coordinates) and is independent of the time, we can say that the distribution of $Q$ is self-similarly expanding at the speed $V_{A0}$. We have verified that our case is just the case of self-similar evolution. In our case, $Q$ denotes any of the variables $\rho$, $P$, $\mbox{\boldmath$v$}$, $\mbox{\boldmath$B$}$, and any combination of these quantities (e.g., magnetic pressure, total pressure, etc.) Hence we can conclude that our result shows the self-similar evolution. 

\placefigure{fig:series1f5}

\placefigure{fig:series2f6}

Figure \ref{fig:pm_vf7} shows the evolution of the velocity field. We can see a self-similar evolutionary character of the velocity field. The upper panel shows the velocity field (red arrows) and the magnetic pressure distribution (color contours) at $t=2449.$, and the lower panel shows the distribution of the same quantities at $t=8818.1$. These two panels look almost the same, except that the spatial scale is different. 

Just after the onset of reconnection, the solution changes form as time proceeds even if it is expressed in the zoom-out coordinates (i.e., $t \mathrel{\hbox{\rlap{\hbox{\lower4pt\hbox{$\sim$}}}\hbox{$<$}}} 1200$). The evolution during this stage depends significantly on the resistivity model. During the period much later from the onset (say, $t \mathrel{\hbox{\rlap{\hbox{\lower4pt\hbox{$\sim$}}}\hbox{$>$}}} 1200$), the evolution gradually settles to the self-similar stage (stationary solution in the zoom-out coordinates). All of the physical quantities shown in figures \ref{fig:profile_rhof8} (density), \ref{fig:profile_pgf9} (gas pressure), \ref{fig:profile_bxf10} ($B_x$; the $x$-component of magnetic field), \ref{fig:profile_vxf11} ($V_x$; $x$-component of velocity) and \ref{fig:profile_vyf12} ($V_y$; the $y$-component of velocity) clearly show such relaxation to the self-similar stage. These figures represent the distribution of physical quantities in the zoom-out coordinates. Note that the scale of the horizontal axis varies in proportion to time. Also note that the location $x$ of the cross section placed parallel to the $y$-axis is also shifted in proportion to time ($x=0.3 V_{A0} t$). This procedure ensures that we take a cross section at the same place in the zoom-out coordinates. We can clearly see that the solution in the zoom-out coordinates is quasi-stationary. This is equivalent to saying that the evolution is self-similar. The reconnection system unlimitedly grows self-similarly once the solution approaches the stationary solution in the zoom-out coordinates. 

\placefigure{fig:pm_vf7}

\placefigure{fig:profile_rhof8}

\placefigure{fig:profile_pgf9}

\placefigure{fig:profile_bxf10}

\placefigure{fig:profile_vxf11}

\placefigure{fig:profile_vyf12}

One may think that the estimated duration of time required for reaching the self-similar stage ($t \sim 1200$) is so much longer than the Alfv\'{e}n transit time scale which is unity in our normalization. We should note that the formation of the slow shock is not caused by the FRW propagation itself, but is caused by the induced inflow. The speed of this inflow is found to be of the order of $10^{-2} V_{A0}$. Hence the inflow transit time scale is of the order of $10^2$ in the normalized time unit. The relaxation time scale of this flow may be estimated to be several times the inflow transit time scale. Thus, the period $t \sim 1200$ is comparable to the inflow relaxation time scale. 

The detailed structure of the self-similar solution at $t=8818$ is shown in figures \ref{fig:det-totf13} and \ref{fig:det-magf14}. The color contours show the distribution of the total pressure (figure \ref{fig:det-totf13}) and magnetic pressure (figure \ref{fig:det-magf14}). The white lines are magnetic field lines. The red arrows show velocity vectors of plasmas. The blue arc denotes FRWF. The velocity field inside the reconnection outflow is intentionally omitted to clarify the velocity structure of the inflow region. 

In figure \ref{fig:det-magf14}, we see the reconnection jet confined by the slow-mode shock along the current sheet ($y \sim 0$, $0 \leq x \mathrel{\hbox{\rlap{\hbox{\lower4pt\hbox{$\sim$}}}\hbox{$<$}}} 2000$). We can also see the plasmoid enclosed by the slow shock ($y \sim 0$, $2000 \mathrel{\hbox{\rlap{\hbox{\lower4pt\hbox{$\sim$}}}\hbox{$<$}}} x \mathrel{\hbox{\rlap{\hbox{\lower4pt\hbox{$\sim$}}}\hbox{$<$}}} 7000$). The magnetic energy is mainly converted at the slow shock. Hence this is ``fast reconnection'' similar to the Petschek model. The contours of the total pressure (see figure \ref{fig:det-totf13}) represent the effect of the fast-mode wave propagation. FRW is produced by the ejection of the reconnection jet. As FRW propagates from the reconnection point ($x=y=0$), the total-pressure decreases in the vicinity of the reconnection point (the blue colored region in figure \ref{fig:det-totf13}). The resultant total pressure gradient induces the inflow toward the reconnection point. The inflow slightly converges as it approaches the current sheet. This convergence is a property of the fast-mode rarefaction-dominated inflow as pointed out by Vasyliunas (1975). These properties (slow shock formation and converging inflow)  suggest that the central region of this reconnection system is indeed very similar to that of the Petschek model. However, we should note the following differences from the original Petschek model. 

The ejected reconnection jet and the plasmoid propagate along the initial current sheet. This strong, dense flow of plasma produces a high total-pressure region (red region in figure \ref{fig:det-totf13}) near the spearhead of the plasmoid by the ``piston'' effect (the fast-mode compression). A vortex-like return-flow gushes out from this high total-pressure region. This return-flow is the property of the evolution process that has not been pointed out by previous works. A fast-mode shock is formed in front of the plasmoid at the interface between the outflow driven by the reconnection and the plasmas in the initial current sheet, but we cannot identify it in figures \ref{fig:det-totf13} and \ref{fig:det-magf14} since the scale of the fast shock is so small as compared to the total system length. To go into details about the structure of the reconnection jet and the plasmoid is beyond the scope of this paper. 

\placefigure{fig:det-totf13}

\placefigure{fig:det-magf14}

\section{Summary \& Discussion}
\label{sec:S-D}

\subsection{Summary}
\label{sec:Sum}

We have presented a new model for describing the evolutionary process of magnetic reconnection: the self-similar evolution model. 

The possibility of this type of evolution has been demonstrated by numerical simulations. An outline of the evolutionary process has been given in section \ref{sec:Sce}. 

We should note that, in this new model, propagation of the fast-mode rarefaction wave (FRW) plays an important role in the self-similar expansion of the reconnection system. FRW propagates almost isotropically at a constant speed $V_{A0}$ (the Alfv\'{e}n speed in the asymptotic region) for the case of low $\beta$ plasmas ($\beta \ll 1$). The only characteristic scale of the system at a later stage (the self-similar evolution stage, see figure \ref{fig:stage3f4}) is the radius of the circular wave front of FRW (FRWF) because the initial current sheet thickness $D$, which is a fixed characteristic scale, is negligible in comparison with the scale of FRWF in this stage. This fact suggests the possibility of a self-similar solution. In fact, the results of our numerical simulation (see figures \ref{fig:series1f5}, \ref{fig:series2f6}) clearly show that the evolution settles to a stationary state if they are expressed in the zoom-out coordinates (see section \ref{sec:Sim}). This is equivalent to saying that such evolution is self-similar. Again note that the self-similar evolution is achieved only at the last stage of evolution ($t \gg D/V_{A0}$) in which the fixed characteristic scale ($D$ or the size of the resistive region) is negligible as compared to the system size (the radius $V_{A0} t$ of FRWF).  Of course, at that stage, the shape and the size of the resistive region, which are artificially given in the model, no longer influences the evolution. 

This new type of reconnection system represents ``fast'' reconnection. As FRW propagates to the uniform region (the asymptotic region $y \gg D$), an inflow is induced by the decrease of the total pressure near the reconnection point. Figure \ref{fig:det-magf14} clearly shows that a pair of slow shocks is formed by this induced flow toward the current sheet. These slow shocks efficiently convert magnetic energy to kinetic or thermal energy of plasma. Of course, the speed of this energy conversion is determined by the slow-mode wave speed, and is independent of the magnetic diffusion speed ($\propto {R_m}^0$). Hence, this energy release is fast one, and we can call it fast reconnection.

\subsection{Why self-similar?}
\label{sec:Why}

In general, self-similar solutions can exist in the systems having no fixed characteristic scale. Our two-dimensional current sheet system however has a fixed characteristic scale: the initial current sheet thickness $D$. 

The typical scale of the entire system can be estimated from the radius of FRWF ($\sim V_{A0} t$). When $t \mathrel{\hbox{\rlap{\hbox{\lower4pt\hbox{$\sim$}}}\hbox{$<$}}} D/V_{A0}$, the evolution is affected by initial conditions, but such effects gradually diminish as the system scale ($V_{A0} t$) exceeds the current sheet thickness ($D$). When $t \gg D/V_{A0}$, the current sheet thickness is negligible comparing with the system scale. In this situation, the system no longer has any characteristic spatial scale than the radius $V_{A0} t$ of FRWF, and the evolution settles to a self-similar solution. In fact, we found self-similar solutions as discussed in section \ref{sec:Num}. Also the size and the shape of the resistive region no longer influence the evolution, because the dimension of the resistive region is negligible as compared to that of the entire system. In fact, we can confirm that the evolution at the later stage is insensitive to the resistivity model (see figure \ref{fig:case_c_pmf15}). 

\placefigure{fig:case_c_pmf15}

It is important to check whether such self-similar solutions are stable or not. We have verified by our numerical simulations that our self-similarly evolving solution is stable over $\sim 10^2$ of the spatial dynamic range. 

The existence of this self-similar evolution stimulates theoretical efforts to find an analytic solution to the problems of self-similar evolution. This will be discussed in a forthcoming paper. 

Strictly speaking, the two-dimensional treatment (resistivity enhancement elongated in the $z$ direction) of this self-similar model may not be plausible for actual cases which will be triggered by a ``point'' resistivity enhancement. In this case, we must treat the evolution as a three-dimensional problem. Differences of such 3-D evolution from the 2-D self-similar evolution model will be treated as a future problem. 

Biernat, Heyn and Semenov (1987) and Semenov et al. (1992) analytically studied the evolutionary process of magnetic reconnection in a similar situation to this paper. They treat the reconnection rate as a free parameter which can be varied arbitrarily. In this meaning, the case they treated is also categorized as spontaneous reconnection. Their analysis gives a general formalism of spontaneous time-varying reconnection. However, in their works, plasma of the inflow region is assumed to be incompressible for analytical convenience. This means that the sound speed is infinitely large even when the Alfv\'{e}n speed is finite. This is equivalent to assuming that the inflow region is filled with extremely high $\beta$ plasmas (note that $\beta \sim$ (sound speed)$^2$/ (Alfv\'{e}n speed)$^2$). Needless to say, this assumption is unsuitable for most cases of astrophysical problems. Contrary to our case, fast-mode waves emitted from the central region instantly propagates to infinity. Although the proper spatial scale determined by the fast-mode wave propagation does not exist in their case,  there is a possibility that the structure formed by slow-mode wave can be self similar, because the propagation speed of the slow-mode wave is finite in their case. Our work is understood as an extension of their works to realistic low $\beta$ astrophysical plasma systems.

\subsection{Adequacy of resistivity model}
\label{sec:Ade}

In our simulation, localization and stationarity of the resistivity are essential assumptions. Since the resistivity cannot be derived from MHD equations, we must apply a certain model for resistivity. In the problem of magnetic reconnection in astrophysical plasmas, so-called anomalous resistivity plays a part in the magnetic diffusion. In order to make a realistic model, the physical process causing the anomalous resistivity is very important, but remains as an open question. 

In addition, as discussed in section \ref{sec:Int}, there is room for discussion concerning two different points of view about influences of the resistivity on the evolution of magnetic reconnection. Even in the case where resistivity does not play crucial roles, there is a prediction that the fast reconnection will never be attained in a system with uniformly distributed resistivity (Biskamp 1986, Scholer 1989, Yokoyama \& Shibata 1994). On the other hand, in the case where resistivity plays crucial roles, the reason why the resistivity can be localized is still unclear. We know neither the origin of resistivity nor influence of resistivity on the reconnection system. 

Our opinion on the resistivity model is that localization and stationarity of the resistivity are essential for the self-similar evolution but the evolution does not depend upon the size or shape of the resistive region nor the physical process causing the resistivity. 

In order to justify our opinion, we must discuss our simulation results with other resistivity models. As an example for a different resistivity model (case B), we have performed the following case (Nitta 1988). We assume a circular resistive region $r \leq D$  in the current sheet, where $r$ is the distance from the origin and $D$ is the initial thickness of the current sheet. In this resistive region, the resistivity is distributed as an exponentially decaying function of $r$. Simulation with this resistivity model also results in self-similar evolution of magnetic reconnection. 

Another example (case C) we have tested has the same shape of the resistive region as the previous case (resistivity uniformly enhanced in the central square region), but the value of the resistivity is five times larger ($\eta=0.5$ thus $R_m=4.9$) than in the previous case ($\eta=0.1$ thus $R_m=24.5$) . The result (see figure \ref{fig:case_c_pmf15}) is very similar to the original case (see figure \ref{fig:series1f5}). The evolution does not depend crucially on the value of resistivity. 

In conclusion, we cannot identify any essential difference among these three cases. Hence, we believe that any model with a localized resistivity  and an infinitely large system size will result in the self-similar evolution. 

Magnetic reconnection needs some processes to keep electric field (reconnection electric filed) along the reconnection line in order to break up the frozen-in condition. In the diffusion region of the plasma system with extremely large magnetic Reynolds number like in astrophysical problems, macroscopic MHD approximation is no longer valid, and the reconnection electric filed should be caused by microscopic processes. Three resistivity models we have examined above are all based on the Ohmic term $\eta J$ where $\eta$ is the resistivity and $J$ is the current density in the diffusion region. In these models, the localized resistivity $\eta$ is supposed to represent anomalous resistivity induced by anomalous collision between particles and waves as a result of micro-instabilities (e.g., the lower hybrid drift [LHD] instability [Huba et al. 1977, Shinohara et al.1998, Shinohara et al. 1999].

However, in actual plasma systems, the reconnection electric field might arise from other terms of the generalized Ohm's law (Biskamp 1997). Several collisionless processes to maintain the reconnection electric field have been proposed (e.g., the Hall current effect [Ma \& Bhattacharjee 1996] or the electron inertia effect [Tanaka 1995], the whistler turbulence effect [Shay \& Drake 1998]). Unfortunately, we do not have a definite strategy for including such reconnection electric field produced by above mentioned microscopic processes in the macroscopic MHD regime. Thus, we adopted a simplified resistivity form in our Ohm's law. 

Of course, the actual resistivity in astrophysical plasma systems cannot be described by simple Ohm's law as adopted in our simulation, but our self-similar evolution only requires a contrivance to keep a finite electric field along the reconnection line. We should emphasize again that the self-similar evolution occurs in the last stage when the system scale becomes much larger than the initial current sheet thickness and the size of the diffusion region. Hence, such evolution can be described by macroscopic MHD, and does not depend upon whether the central resistivity is caused by collisional or collisionless processes. 

In the case of evolution in free space, boundary conditions that are crucial to the ``driven reconnection'' models obviously do not play important roles. Hence the resistivity should control the evolution any way. However, we do not go into details of the nature of resistivity itself in this paper, and adopt a simple model of localized and stationary resistivity. The dependence of the self-similar evolution on various resistivity models will be discussed in our forthcoming paper in detail.

\subsection{Comparison with previous stationary models}
\label{sec:Com}

There are several theoretical models for steady state magnetic reconnection. We compare our self-similar evolution model with these previous models. Our discussion is focused only on the fast reconnection, because the very quick energy conversion frequently observed in astrophysical phenomena suggests that fast reconnection should be considered as the responsible mechanism. 

The Petschek model (Petschek 1964) is characterized by a pair of slow shocks and fast-mode rarefaction in the inflow region. This fast-mode rarefaction wave is produced at the central reconnection region. As a result of the fast-mode rarefaction, the gradient of the magnetic field strength near the neutral point decreases due to the bending of magnetic field lines (Vasyliunas 1975). This process limits the diffusion speed, and hence the reconnection rate. 

The Sonnerup model (Sonnerup 1970) is developed from the Petschek model. This model is characterized by a pair of slow shocks and the hybrid of fast-mode and slow-mode rarefaction. The fast-mode rarefaction wave is produced at the central region as in the Petschek model, but the slow-mode rarefaction wave is injected from the boundary. Because of the hybrid nature of rarefaction, the gradient of the magnetic field strength near the neutral point does not decrease as in the Petschek model. Therefore the Sonnerup model can attain the maximum reconnection rate possible for magnetic energy converters (Priest \& Forbes 1986). 

Our self-similar evolution is never influenced by the boundary conditions, so the Sonnerup-type hybrid rarefaction does not take place at all in our case. The fast-mode rarefaction dominates in the vicinity of the diffusion region. As discussed in section \ref{sec:Sim}, the central region of this self-similarly evolving system is of the Petschek-type. Therefore, the reconnection rate will be limited in a way similar to the original Petschek model. This point will be clarified in our following papers. 

Far from the central region, there is a region in which the fast-mode compression takes place by the piston effect of the plasmoid. This fast-mode compression causes the vortex-like return flow (see figure \ref{fig:det-totf13}). We can consider that this self-similar evolution includes the Petschek model as the inner solution. A combined feature of the original Petschek model and the vortex-like flow properly characterizes the evolutionary process, and the entire system unlimitedly expands self-similarly.

\subsection{Applications to flares}
\label{sec:App}

This study has been motivated by the fact that actual reconnection has a very wide spatial dynamic range of evolution; in the case of geomagnetospheric substorms or the solar flares; the dynamic range is $10^4-10^7$. Therefore, the early stage of the evolution of magnetic reconnection can be approximated as the evolution into free space without any influence of boundary conditions. This approximation is justified as long as the spatial scale of the reconnection system is much smaller than that of the entire system (e.g., the radius of curvature of the initial magnetic field lines). In the case of large solar flares, the typical radius of curvature is $\sim 10^{7-8}$ m and the typical Alfv\'{e}n speed is $\sim 10^6$ m s$^{-1}$. Hence we can predict from our self-similar model that during the interval $\sim 10^{7-8}/10^6 \sim 10^{1-2}$ s from the onset of magnetic reconnection, the evolution can be approximated by our self-similar evolution model, and the total power, integrated over the entire system, of the energy conversion increases in proportion to time. The authors expect that coming observations by Solar B will detect such phenomena. 

Another property of our self-similar model is the characteristic structure of the inflow region. A vortex-like flow around the head of the plasmoid characterizes the evolutionary process (see figure \ref{fig:det-totf13}). We may be able to verify our evolutionary model by detecting this kind of vortex flow. 

It has recently been found that solar flares (and similar arcade formation events) are often associated with depletion of ambient plasma density, called dimming, during their main energy release phase (Tsuneta 1996, Sterling and Hudson 1997). Tsuneta (1996) attributed this to the result of reconnection inflow (see also Yokoyama and Shibata 1997), while Sterling and Hudson (1997)  interpreted this as being due to plasma ejection from the system. In the case of a large cusp-shaped flare (Tsuneta 1996), the time scale of the ``dimming'' is 10-20 minutes just at the onset of the flare, which is about 7-14 $t_A$ (where $t_A$ is the Alfv\'{e}n transit time), and the ratio of depleted density to the initial density is $\sim$ 0.2 during this period. Let us examine the possibility that the fast rarefaction wave associated with reconnection inflow can account for the observed "dimming" at the onset of flares. In figure \ref{fig:dimmingf16}, predicted X-ray image (in relative measure) of the reconnection system is shown. This figure is drawn from the result of case A. Temperature range is assumed as $10^7-10^8$ K, which is plausible for solar flares. We can clearly find that a round dark region around the reconnection point expands as the fast mode rarefaction wave propagates isotropically. Our simulation results shown in figure \ref{fig:dimmingf16} show that the ratio of the depleted density to the initial density in the rarefaction wave is of order of 0.1-0.2, consistent with observations (see figure \ref{fig:profile_rhof8}). 

In actual reconnection systems, the spatial scale (say $L$) of the region in which magnetic flux is piled up should be finite. This region is an energy reservoir of magnetic reconnection. Although the time scale of strict self-similar evolution in the bounded system ($L=finite$) is the Alfv\'{e}n time ($t_A \equiv L/V_{A0}$), our self-similar evolution  (especially near the reconnection point) may roughly hold more than the Alfv\'{e}n time (up to a few to 10 $t_A$; see figure \ref{fig:rec_ratef17}) even if the system is not in an exact free space. The solid line of figure \ref{fig:rec_ratef17} shows the time variation of the electric field ($\eta |\mbox{\boldmath$J$}|$) at the reconnection point ($x=y=0$) for the typical case (case A). The intensity of the electric field is a measure of the reconnection rate of the system. The dashed line (case D) shows the reconnection rate for the case where the lower half region ($y \leq 3000$) is filled with uniformly magnetized plasma, while the upper half region ($y>3000$) is unmagnetized. In case D, the fast mode wave is reflected at the boundary at $y=3000 \  (t=3000)$, and returns to the resistive region at $t=6000$. Hence, the physical difference between these cases (case A and D) will appear after $t=6000$. Figure \ref{fig:rec_ratef17} shows that the difference is only a few percent, and moreover, we cannot find any significant difference between these two cases even after the reflected wave returns to the current sheet ($t \geq 6000$). 

Consequently, basic characteristics (density and time scale) of fast-mode rarefaction wave in a self-similar evolution of fast reconnection seem to account for observations of ``dimming'' associated with solar flares.

\placefigure{fig:dimmingf16}

\placefigure{fig:rec_ratef17}


\acknowledgments

We are grateful to Drs. T. Yokoyama at Nobeyama Radio Observatory and T. Kudoh at National Astronomical Observatory of Japan for useful comments and fruitful discussion. The numerical computations were carried out on VPP300/16R and VX/4R at the National Astronomical Observatory of Japan.

\figcaption[sim_boxf1.eps]{
The simulation box and the initial state. We set the $x$ axis of our Cartesian coordinates along the current sheet and the $y$ axis perpendicular to the current sheet. We treat a 2-D MHD problem and assume uniformity in the $z$-direction. An artificially imposed resistive region is placed near the origin, and the resistivity is held constant (the resistivity $\eta=0.1$ in our dimensionless unit in the region $0 \leq x \leq 2D$, $0 \leq y \leq 2D$, and $\eta=0$ outside). The initial state is the Harris solution, so physical quantities in the asymptotic region far outside the current sheet are uniform. 
\label{fig:sim_boxf1}
}

\figcaption[stage1f2.eps]
{
Schematic scenario of the evolutionary process. \\
1): Resistive stage ($t < D/V_{A0}$). When resistivity is locally enhanced in the current sheet, the magnetic field lines begin to reconnect, and a flow filed is induced as in the Sweet-Parker model or resistive tearing instability. Ejection of bipolar plasma flow (reconnection jets) causes decrease of the total-pressure around the reconnection point, and the fast-mode rarefaction wave (FRW) is emitted from the diffusion region. 
\label{fig:stage1f2}
}

\figcaption[stage2f3.eps]
{
2): Induction of inflow ($t \mathrel{\hbox{\rlap{\hbox{\lower4pt\hbox{$\sim$}}}\hbox{$>$}}} D/V_{A0}$). When the fast-mode rarefaction wave front (FRWF) propagates into the asymptotic region (the uniform region at $y \gg D$), the difference of the total pressure between the diffusion region and FRWF induces the inflow. This inflow is slightly converging. This feature is characteristic to the fast-mode rarefaction dominated inflow. 
\label{fig:stage2f3}
}

\figcaption[stage3f4.eps]
{
3): Self-similar evolution stage ($t \gg D/V_{A0}$). When the inflow sufficiently evolves, the inflow speed exceeds the slow mode propagation speed, and a pair of slow-mode shocks is formed along the current sheet. Once this show shock is formed, the energy conversion by the reconnection drastically proceeds (the fast reconnection). We should remark that this fast reconnection system continues to expand self-similarly and unlimitedly as FRWF propagates. 
\label{fig:stage3f4}
}

\figcaption[series1f5.eps]
{
A sequence of six snapshots representing the evolutionary process of magnetic reconnection. The color contours denote the magnetic pressure distribution. The blue arc denotes the wave front of the fast-mode rarefaction wave (FRWF). The system self-similarly evolves after $t \sim 1224.7$ (case A). 
\label{fig:series1f5}
}

\figcaption[series2f6.eps]
{
The evolution subsequent to the sequence in figure \ref{fig:series1f5}. The mesh size for this figure is rougher than that of figure \ref{fig:series1f5}. We can clearly see that the self-similar evolution continues over a large spatial dynamic range (case A'). 
\label{fig:series2f6}
}

\figcaption[pm_vf7.eps]
{
Velocity distribution in the inflow region at different times ($t=2449$ and $t=8818.1$). The velocity vectors are shown by red arrows on the background color contours of the magnetic pressure.  We cannot see any essential difference between these two panels except for the spatial scaling. This indicates self-similarity of evolution of the velocity field. The velocity filed inside the reconnection outflow is intentionally omitted to clarify the velocity in the inflow region, because it is much larger than the inflow velocity (case A). 
\label{fig:pm_vf7}
}

\figcaption[profile_rhof8.eps]
{
Density profile along the $y$ axis in the zoom-out coordinates ($x$ coordinate expands in proportion to time). The profile gradually settles to a stationary state. We can roughly estimate the relaxation time scale as $t \sim 1200$. The depleted density in the rarefaction wave is of the order of 0.1-0.2 times the initial density (case A).
\label{fig:profile_rhof8}
}

\figcaption[profile_pgf9.eps]
{
Gas-pressure profile along the $y$ axis in the zoom-out coordinates. Obvious step-like feature near $x \sim 0.009$ is the Petschek-type slow shock (case A). 
\label{fig:profile_pgf9}
}

\figcaption[profile_bxf10.eps]
{
Magnetic field ($x$-component) profile along the $y$ axis in the zoom-out coordinates. Obvious step-like feature near $x \sim 0.009$ is the Petschek-type slow shock (case A). 
\label{fig:profile_bxf10}
}

\figcaption[profile_vxf11.eps]
{
Velocity field ($x$-component) profile along the $y$ axis in the zoom-out coordinates. Obvious step-like feature near $x \sim 0.009$ is the Petschek-type slow shock (case A). 
\label{fig:profile_vxf11}
}

\figcaption[profile_vyf12.eps]
{
Velocity field ($y$-component) profile along the $y$ axis in the zoom-out coordinates. Obvious step-like feature near $x \sim 0.009$ is the Petschek-type slow shock. These profiles of physical quantities (figures \ref{fig:profile_rhof8}, \ref{fig:profile_pgf9}, \ref{fig:profile_bxf10}, \ref{fig:profile_vxf11} and \ref{fig:profile_vyf12}) clearly show that the solution is quasi-stationary in the zoom-out coordinate. This is equivalent to the presence of the self-similar evolution (case A). 
\label{fig:profile_vyf12}
}

\figcaption[det-totf13.eps]
{
Detailed structure of the self-similar solution. The total-pressure distribution is shown by color contours. The effects of the fast-mode wave are apparent in this figure. The effect of fast-mode rarefaction wave (FRW) is clearly seen near the reconnection point. The piston effect (the fast-mode compression) takes place near the spearhead of the plasmoid. The inner structure near the reconnection point is quite similar to the Petschek model (case A). \label{fig:det-totf13}
}

\figcaption[det-magf14.eps]
{
Detailed structure of the self-similar solution. The magnetic pressure distribution is shown as color contours. The effects of the slow-mode wave are apparent in comparison with figure \ref{fig:det-totf13}. A pair of slow-mode shocks, the reconnection jet, and the plasmoid are clearly shown (case A). 
\label{fig:det-magf14}
}

\figcaption[case_c_pmf15.eps]
{
The same as in figure \ref{fig:series1f5}, but for the case of larger resistivity (five times larger than that in the case of figure \ref{fig:series1f5}; $\eta=0.5$, $R_m=4.9$). We find no essential difference from the previous case. This result suggests that the self-similar evolution does not sensitively depend upon details of the resistivity model (case C). 
\label{fig:case_c_pmf15}
}

\figcaption[dimmingf16.eps]
{
X-ray flux distribution predicted by our model in relative measure. This figure shows the evolution of a virtual X-ray image in the fixed coordinates. The contours indicate the distribution of the emission measure $f \propto n^2 T^{1/2}$ which shows X-ray flux distribution in the assumed temperature range of $10^7-10^8$ K , where $n$ is the particle number density and $T$ the temperature. Our self-similar model may be applied to the observation of dimming (case A). 
\label{fig:dimmingf16}
}

\figcaption[rec_ratef17.eps]
{
Evolution of the reconnection rate. The solid line shows the time variation of the electric field ($\eta |\mbox{\boldmath$J$}|$) at the reconnection point ($x=y=0$) in the case shown in figure \ref{fig:series1f5} and \ref{fig:series2f6}. This quantity may be taken as the reconnection rate of the system (case A). In the same way, the dashed line shows it for the case the lower half region ($y \leq 3000$) is filled with uniformly magnetized plasma, while the upper half region ($y>3000$) is unmagnetized. We cannot find any significant difference between these two cases even after the reflected wave from the interface between magnetized and unmagnetized regions returns to the current sheet ($t \geq 6000$) (case D). 
\label{fig:rec_ratef17}
}


\begin{thebibliography}{99}
\baselineskip=1.0pc

\bibitem[Biernat 1987]{BHS}
Biernat, H.K., Heyn, M. \& Semenov, V.S., 1987, JGR, 92, A4, 3392

\bibitem[Biskamp 1986]{Bis1}
Biskamp, D. 1986, Phys. Fluids, 29, 1520

\bibitem[Biskamp 1986]{Bis2}
Biskamp, D., 1997, Phys. Plasmas, 4 (5), 1964

\bibitem[Huba et al. 1977]{HGP}
Huba, J. D., Gladd, N. T. \& Papadopoulos, K., 1977, GRL, 4, 125

\bibitem[Ma \& Bhattacharjee 1996]{M-B}
Ma, Z. W. \& Bhattacharjee, A., 1996, GRL, 23, 1673

\bibitem[Nitta 1988]{Nit88}
Nitta, S. 1988, Master thesis, Nagoya University (in Japanese)

\bibitem[Parker 1963]{Par}
Parker, E. N. 1963, ApJ Suppl. Ser., 8,177

\bibitem[Petschek 1964]{Pet}
Petschek, H. E. 1964, NASA Spec. Publ., 50, AAS-NASA Symposium on Physics of Solar Flares, 425

\bibitem[Priest \& Forbes 1986]{P-F}
Priest, E. R. \& Forbes, T. G. 1986, J. Geophys.Res, 91, 5579

\bibitem[Sato \& Hayashi 1979]{S-H}
Sato, T. \& Hayashi, T. 1979, Phys. Fluids, 22, 1189

\bibitem[Scholer 1989]{Sch}
Scholer, M. 1989, JGR, 94, 8805

\bibitem[Semenov 1992]{SKLRHB}
Semenov, V. S., Kubyshkin, I. V., Lebedeva, V. V., Rijnbeek, R. P., Heyn, M. \& Biernat, H. K., 1992, Planet. Space Sci., 40, 63

\bibitem[Shay \& Drake 1998]{S-D}
Shay, M. A. \& Drake, J. F., 1998, GRL, 25, 3759

\bibitem[Shinohara et al. 1998]{SNFTMTY1998}
Shinohara, I., Nagai, T., Fujimoto, M., Terasawa, T., Mukai, T., Tsuruda, K. \& Yamamoto, T., 1998, JGR, 103, 20365 

\bibitem[Shinohara \& Hoshino 1999]{S-H99}
Shinohara, I. \& Hoshino, M., 1999, Adv. in Space Res., 24, 43

\bibitem[Sonnerup 1970]{Son}
Sonnerup, B. U. \"{O}, J. Plasma Phys. 1970, 4, 161

\bibitem[S-H 1997]{Ste}
Sterling, A. C. \& Hudson, H. S. 1997, ApJ 491 L55

\bibitem[Sweet 1958]{Swe}
Sweet, P. A. 1958, The neutral point theory of solar flares, in Electro-magnetic Phenomena in Cosmical Physics, ed. B.Lehnert (Cambridge University Press, London), 123

\bibitem[Tajima \& Shibata 1997]{T-S}
Tajima, T. \& Shibata, K. 1997, Plasma Astrophysics (Reading, Massachusetts: Addison-Wesley), 223

\bibitem[Tanaka 1995]{T95}
Tanaka, M., 1995, Phys. Plasmas, 2, 2920

\bibitem[Tsuda \& Ugai 1977]{T-U}
Tsuda, T. \& Ugai, M. 1977, J. Plasma Phys., 18, 451

\bibitem[Tsuneta 1996]{Tsu}
Tsuneta, S. 1996, ApJ, 456, 840 [erratum see ApJ, 464, 1055]

\bibitem[Ugai \& Tsuda 1977]{U-T1}
Ugai, M. \& Tsuda, T. 1977, J. Plasma Phys., 17, 337

\bibitem[Ugai \& Tsuda 1979]{U-T2}
Ugai, M. \& Tsuda, T. 1979, J. Plasma Phys., 22, 1

\bibitem[Ugai 1999]{Uga2}
Ugai, M. 1999, Phys. Plasmas, 6, 1522

\bibitem[Vasyliunas 1975]{Vas}
Vasyliunas, V. M. 1975, Rev. Geophys., 13, 303

\bibitem[Yokoyama \& Shibata 1994]{Y-S}
Yokoyama, T. \& Shibata, K. 1994, ApJ, 436, L197

\bibitem[Yokoyama \& Shibata 1997]{Y-S2}
Yokoyama, T. \& Shibata, K. 1997,
Fifth SOHO Workshop: The Corona and Solar Wind Near
Minimum Activity. held at Institute of Theoretical Astrophysics.
 University of Oslo, Norway, 17-20 June, 1997.
Edited by A. Wilson, European Space Agency, p.745

\end{thebibliography}
\end{document}